\documentclass[12pt]{article}
\usepackage{epsfig}
\usepackage{graphicx}

\def\beq{\begin{equation}}
\def\eeq#1{\label{#1}\end{equation}}
\def\eeqn{\end{equation}}

\def\beqa{\begin{eqnarray}}
\def\eeqa#1{\label{#1}\end{eqnarray}}
\def\eeqan{\end{eqnarray}}

\let\bar=\overbar

\def\Dslash{\not{\hbox{\kern-4pt $D$}}}
\def\dslash{\not{\hbox{\kern-2pt $\del$}}}

\def\msb{{\bar{\ssstyle M \kern -1pt S}}}

\usepackage{fancyhdr,graphicx}
\def\Title#1{\begin{center} {\Large {\bf #1} } \end{center}}

\begin{document}

\Title{Braking index of isolated uniformly rotating magnetized pulsars}

\bigskip\bigskip

%+\addcontentsline{toc}{chapter}{{\it D. Blaschke}}
%+\label{BlaschkeDavid}

\begin{raggedright}

{\it 
Oliver Hamil$^{1}$~~Jirina Stone$^{1,2}$~~Martin Urbanec$^{3}$~~and Gabriela Urbancova$^{3,4}$\\
%\thanks{\tt Email:}
\bigskip
$^{1}$Department of Physics and Astronomy, University of Tenessee, Knoxville, Tenessee 37996, USA\\
\bigskip
$^{2}$Department of Physics, Oxford University, Oxford, United Kingdom\\
\bigskip
$^{3}$Institute of Physics, Faculty of Philosophy and Sciences, Silesian University in Opava, CZ 74601 Opava, Czech Republic\\
\bigskip
$^{4}$Speaker who presented the work described herein\\
}

\end{raggedright}

\section{Introduction}

The slowing down of rotating neutron stars has been observed and modeled for decades. The simplest models relate the loss the kinetic rotational energy of the star to the emission of magnetic radiation from a rotating dipolar magnetic field (MDR), attached to the star \cite{pacini1967, pacini1968, gold1968, gold1969,goldwire1969}. The calculated energy loss by a rotating pulsar is assumed proportional to a model dependent power of $\Omega$. This relation leads to the power law $\dot{\Omega}$ = -K $\Omega^{\rm n}$ where $n$ is called the braking index. The value of $n$ can be, in principle, determined from observation of higher-order frequency derivatives related to n by \cite{lyne2015}
\begin{eqnarray}
n& = & {{{\Omega} \ddot{\Omega}} \over {{\dot{\Omega}}^2}} \\
%n(2n-1)& =& {{{\Omega}^2 \dddot{\Omega}} \over {{\dot{\Omega}}^3}}.
\end{eqnarray}

When the star is taken as a magnetized sphere, rotating \textit{in vacuum}, with a constant moment of inerta (MoI) and a constant magnetic dipole moment, missaligned at a fixed angle to its axis of rotation, $n$ is equal to 3 (for derivation see Section~\ref{subsec:stat}). 

Extraction of the rotational frequency and its time derivatives from observation involves a detailed analysis of the time evolution of the pulses, and of the spectra and luminosity of radiation from the related nebulae in a wide range of wavelengths. Although data on many pulsars are available in the literature, there are only eight pulsars generally accepted to yield reliable data on the pulsar's spin-down (see Table.~\ref{tab1}, recent compilation \cite{magalhaes2012} and Refs. therein). The third derivative is known only for the Crab pulsar \cite{lyne1988}, and PSR B1509-58 \cite{kaspi1994}.

Examination of Table~\ref{tab1} shows that $n = 3$ does not agree with observation. There have been many attempts to extend/modify the basics of the MDR model.  These include consideration of magnetic field activity (e.g. \cite{livingstone2011,blanford1988, melatos1997, lyne2004,harding1999,kramer2006, lyne2010}), superfluidity and superconductivity of the matter within pulsars (e.g. \cite{sed1998, ho2012, page2014}), and modifications of the power law and related quantities (e.g. \cite{johnston1999, magalhaes2012}). Time dependence of the constants in the MDR model has also been considered \cite{blanford1988, contopoulos2006, zhang2012, gourgouliatos2014}. In particular, time evolution of the inclination angle between spin and magnetic dipole axes as been recently addressed \cite {lyne2015,lyne2013}. However, there is no model currently available which would yield, consistently, the typical spread of values of $n$ as illustrated in Table~\ref{tab1}.

%%%%%%%%%%%%%%%%%%%%%%%%%%%%%%%%%%%%%%%%%%%%%%%%%%%%%%%%%%%%%%%%%%%%%%%%%
%%
%%   use this format if you need to include a LaTeX table into your paper
%%
\begin{table}
\centering{}%
\begin{tabular}{|c|c|c|c|c|}
\hline 
PSR                           &       Frequency          &  $n$                             & Ref.  \\
                                     &            (Hz)               &                                 &     \\ \hline
 B1509$-$58              &     6.633598804    &    2.839$\pm$0.001          &  \cite{livingstone2007}  \\
 J1119$-$6127          &     2.4512027814   &   2.684$\pm$0.002          & \cite{waltevrede2011}   \\
 J1846$-$0258          &     3.062118502     &   2.65$\pm$0.1                &\cite{livingstone2007}     \\
                                &                             &   2.16$\pm$0.13                             &\cite{livingstone2011} \\
 B0531+21 (Crab)      &    30.22543701      &   2.51$\pm$0.01               &  \cite{lyne1993} \\
 B0540$-$69             &    19.8344965        &   2.140$\pm$0.009            & \cite{livingstone2007,boyd1995}  \\
 J1833$-$1034         &    16.15935711      &   1.8569$\pm$0.001       & \cite{roy2012}   \\
 B0833$-$45 (Vela)    &    11.2                   &   1.4$\pm$0.2                   & \cite{lyne1996}   \\
 J1734$-$3333          &      0.855182765    &    0.9$\pm$0.2                &\cite{espinoza2011}   \\  \hline 
\end{tabular}\medskip{}
 \caption{Selected pulsars adopted from \cite{magalhaes2012,espinoza2011,lyne2015}. 
\label{tab1}}
\end{table}
%%%%%%%%%%%%%%%%%%%%%%%%%%%%%%%%%%%%%%%%%%%%%%%%%%%%%%%%%%%%%%%%%%%%%%%%%%%

In this work we focus on determination of the maximum deviation of the braking index from the value $n = 3$ by introducing two modifications of the simple MDR model: frequency dependence of MoI, related to the change of shape of a deformable star due to rotation, and the macroscopic effect of superfluidity of the pulsar core. The correction to the expression the braking index arising from these modifications following Glendenning \cite{glendenning} is derived, and included in the calculation, using four realistic Equations of State (EoS) over range of baryonic mass (M$_{\rm B}$) . We study the relation between the softness of the EoS and the rate of change of the braking index as a function of frequency and the M$_{\rm B}$). The four EoS's were also used to obtain mass density profiles of the pulsars needed to determine the transition region between the crust and core. These results were utilized in simulation of an effect of superfluid conditions which eliminate the angular momentum exchange at the threshold between the crust and core. The calculation is performed over a full range of frequencies of the pulsar from zero to the Kepler frequency and a range of M$_{\rm B}$ from 1.0 to 2.2 M$_\odot$, representing the gravitational mass range from about 0.8 to 2.0 M$_\odot$.

\section{Simple MDR Model}
\label{subsec:stat}
The total energy loss by a rotating magnetized sphere can be expressed in terms of the time derivative of the radiated energy as \cite{pacini1968, glendenning, lyne2015}
\begin{eqnarray}
{dE \over dt} &=& -{2\over3}\mu^{\rm 2} {\Omega}^4 {\sin^2{\alpha}},
\label{eq:2}
\end{eqnarray}
where $\mu$ is the magnetic dipole moment of the pulsar, $\mu$ = B R$^3$. $R$ is the radial coordinate of a surface point with the surface magnetic field strength $B$, $\Omega$ is the rotational frequency, and $\alpha$ is the angle of inclination between the dipole moment and the axis of rotation \cite{glendenning}. 

Substituting the kinetic energy of a rotating body, dependent on the MoI $I$,
\begin{eqnarray}
E &=& {1\over2}I\Omega^2,
\label{eq:2.1}
\end{eqnarray}
into (\ref{eq:2}) yields
\begin{eqnarray}
{d\over dt}\left({1\over2}I\Omega^2\right) &=& -{2\over3}\mu^2 {\Omega}^4 {\sin^2{\alpha}}.
\label{eq:2.3}
\end{eqnarray}
Assuming constant MoI, $dI/dt$ = 0, we get
\begin{eqnarray}
%{1\over2}I2\Omega \dot{\Omega} &=& -{2\over3}\mu {\Omega}^4 {\sin^2{\alpha}}\\
\dot{\Omega} &=& -{2\over3}{{\mu^{\rm 2}}\over{I}} {\Omega}^3 {\sin^2{\alpha}}.
\label{eq:2.4}
\end{eqnarray}
Setting $K = {2\over3}{{\mu^2}\over{I}}  {\sin^2{\alpha}}$ in (\ref{eq:2.4}) and taking $\mu$ and $\alpha$ constant leads to the commonly used braking  power law describing the pulsar spin-down due to dipole radiation:
\begin{eqnarray}
\dot{\Omega} &=& -K\Omega^3.
\label{eq:2.5}
\end{eqnarray}
Differentiating (\ref{eq:2.5}) with respect to time
\begin{eqnarray}
\ddot{\Omega} &=& -3K\Omega^2\dot{\Omega},
\label{eq:2.6}
\end{eqnarray}
and combining (\ref{eq:2.5}) and (\ref{eq:2.6}) to eliminate $K$ we get the value of the braking index $n$
%And finally, substituting Eqs.\ref{eq:2.5} and \ref{eq:2.6} into Eq.\ref{eq:1}, we get the canonical neutron star braking index value for a spherical pulsar with constant $R$,{\bf$B$},$I$, and $\alpha$;
\begin{eqnarray}
n =  {{{\Omega} \ddot{\Omega}} \over {{\dot{\Omega}}^2}} &=& 3.
\label{eq:2.7}
\end{eqnarray}
 
\subsection{MDR model with frequency dependent MoI}
\label{subsec:dyn}
The simple MDR value $n=3$ (\ref{eq:2.7}) is derived taking the  $I$, $\mu$ and $\alpha$ as independent of frequency and constant in time. However, in reality the MoI of rotating pulsars changes with frequency and, consequently with time. \cite{gle1997, glendenning}. The equlibrium state of a rotating pulsar includes the effect of centrifugal forces, acting against gravity. The shape of the pulsar is ellipsoidal with decrease (increase) in radius along the equatorial (polar) direction with respect to the rotation axis as pulsar spins down. Thus the MoI, and, consequently, the braking index, are both frequency dependent.

It is convenient to re-write (\ref{eq:2.3}) as
\begin{eqnarray}
{d \over dt}\left({1\over2}I\Omega^2\right) &=& -C\Omega^4,
\label{eq:3}
\end{eqnarray}
\noindent
where $C= {2\over3}\mu^2 sin^2\alpha  $.
Assuming this time that $dI/dt$ is non-zero, differentiation of (\ref{eq:3}) with respect to time gives
\begin{eqnarray}
2I\dot{\Omega} + \Omega \dot{I} = -2C\Omega^3.
\label{eq:3.1}
\end{eqnarray}
Differentiating once more gives
\begin{eqnarray}
2I\ddot{\Omega} + 2\dot{\Omega}\dot{I} + \dot{\Omega}\dot{I} + \Omega\ddot{I} = - 6C\Omega^{2}\dot{\Omega}.
\label{eq:3.2}
\end{eqnarray}
Using the chain rule we can write $\dot{I}$ in terms of $\dot{\Omega}$
\begin{eqnarray}
{dI \over dt} = {d\Omega \over dt}{dI \over d\Omega},
\end{eqnarray}
and obtain
\begin{eqnarray}
\dot{I} = I'\dot{\Omega}\\ \ddot{I} = {\dot{\Omega}}^2I'' + I'\ddot{\Omega},
\end{eqnarray}
where the primed notation represents the derivative with respect to $\Omega$.

Substituting the identities shown above into (\ref{eq:3.1}) and (\ref{eq:3.2}), we get the following relations for $\dot{\Omega}$ and $\ddot{\Omega}$,
\begin{eqnarray}
\dot{\Omega} &=& -{2C\Omega^2} \over {(2I + \Omega I')} \\
\ddot{\Omega} &=& {{-6C\dot{\Omega}\Omega^{2} -\dot{\Omega}^2(3I'+\Omega I'')} \over {(2I + \Omega I')}}.
\end{eqnarray}
After some algebra it is easy to show that the expression of the braking index as a function of angular velocity reads
\begin{eqnarray}
n(\Omega) &=&{{\Omega \ddot{\Omega}} \over \dot{\Omega}^2} = 3 -{{(3\Omega I' + \Omega^2 I'')} \over {(2I + \Omega I')}}.
\label{eq:3.3}
\end{eqnarray}

We note that the magnetic dipole moment of the non-spherical pulsar may in principle also change with frequency. Estimation of this effect would require knowledge of the origin and distribution of the dipole moment, which is lacking. We therefore ignore such change here and restrict ourselves to analysis of the two effects described.
 
\section{Calculation Method} \label{sec:calc}
Previous modeling of the braking index using the simple MDR model with constant MoI assumed a pulsar with $1.4 M_{\odot}$ gravitational mass and a radius $\sim$ 10 km. In this work, which includes frequency dependent MoI and varying M$_{\rm B}$, we solve the equations of motion of rotating stars with realistic EoS using two different numerical methods. 
 
\subsection{The codes}
The PRNS9 code, developed by Weber \cite{book:weber, weber_private}, is based on a perturbative approach to the equations of motion of  slowly rotating near-spherical objects \cite{hartle1968, hartle1973}. To ensure the reliability of the PRNS9 code results, we also used the RPN code. This code by Rodrigo Negreiros \cite{negreiros} is based on a publically available algorithm, RNS, developed by Stergioulas and Friedman \cite{RNS}. The equations of motion are derived directly from Einstein's equations, following the  Cook, Shapiro and Teukolsky approach \cite{CST}, described in detail in \cite{komatsu1989}. Both codes are applicable to rotating stars with all frequencies up to the Kepler limit.

\begin{figure}[htb]
 \includegraphics[width=0.6\textwidth]{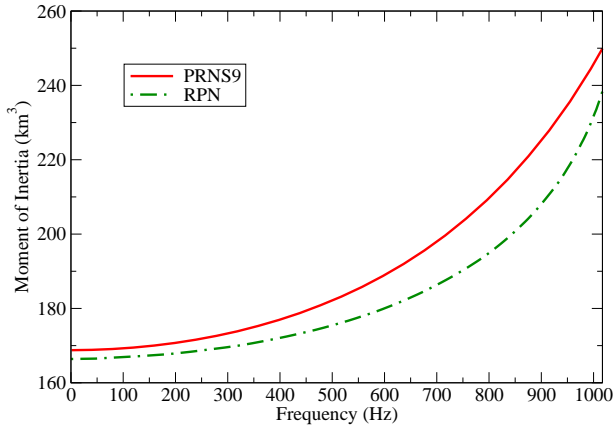}
 \caption{MoI as a function of frequency for a pulsar with M$_{\rm B}$  = 2.0~$M_\odot$  as calculated with both, RPN and PRNS9 numerical codes.}
\label{fig1}
\end{figure}

A comparison of the results of the two codes is demonstrated in Figure~\ref{fig1} which shows MoI as a function of frequency for a pulsar with the QMC700 EoS and M$_{\rm B}$ = 2.0 M$_\odot$ (see Section~\ref{subsec:eos}). They differ most, but by less than 10\%, as the Kepler frequency is approached.  The small difference at near zero frequency (about 1.25\%), due to the difference in behavior of the two low density EoS's (see Section~\ref{subsec:eos} ), is negligible in the context of calculating neutron star macro-properties.  

\subsection{The Equation of State} \label{subsec:eos}

An essential input to the calculation of macroscopic properties of rotating neutron stars is the EoS. The EoS is constructed for two physically different regimes, the high density core and the relatively low density crust.

\begin{figure}[htb]
 \includegraphics[width=0.6\textwidth]{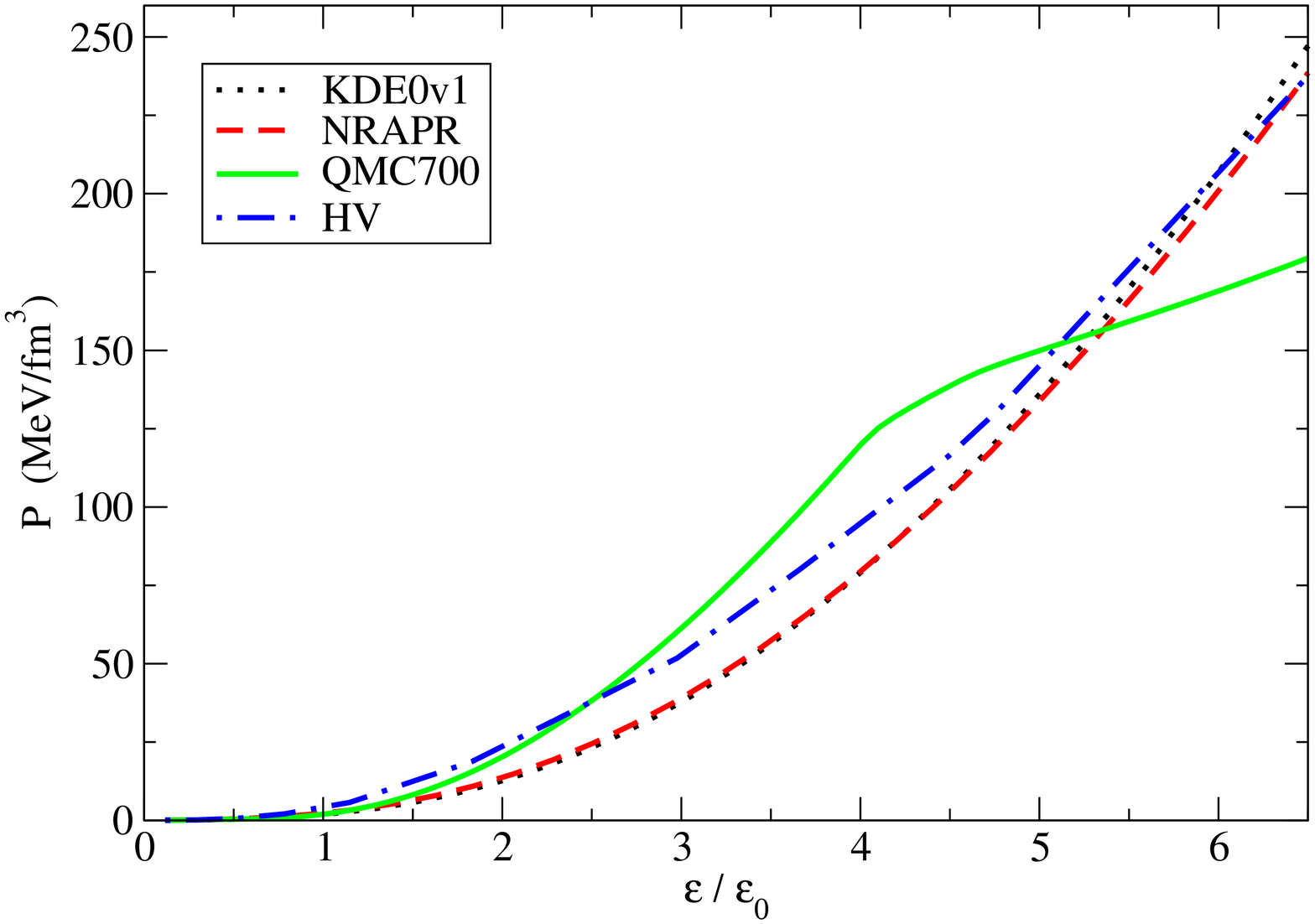}
 \caption{Pressure vs. energy density $\epsilon$ (in units of the energy density of symmetric nuclear matter at saturation $\epsilon_{\rm 0}$) as predicted by the four EoS's used in this work.}
\label{fig2}
\end{figure}

The microscopic composition of high density matter in the cores of neutron stars is not well understood. We have chosen two EoS's, which assume that the core is made only of nucleons, KDE0v1 \cite{agrawal2005} and NRAPR \cite{steiner2005}. These EoS were selected by Dutra et al. \cite{dutra2013} as being among the very few which satisfied an extensive set of experimental and observational constraints on properties of high density matter. In addition, we use two more realistic EoS which include in the core the heavy strange baryons (hyperons) as well as nucleons. The QMC700 EoS has been derived in the framework of the Quark-Meson-Coupling (QMC) model \cite{guichon2006, stone2007} and the Hartree V (HV) EoS \cite{weber1989} is based on a relativistic mean-field theory of nuclear forces.  The maximum mass of a static star, calculated using the Tolman-Oppenheimer-Volkoff (TOV) equation, is 1.96, 1.93, 1.98 and 1.98 M$_\odot$ for KDE0v1, NRAPR, QMC700 and HV, respectively, which is close to the gravitational mass of the  heaviest known neutron stars \cite{demorest2010, antoniadis2013}.  The EoS's are illustrated in Figure~\ref{fig2} which shows pressure as a function of energy density $\epsilon$ in units of nuclear saturation energy density $\epsilon_0$ = 140 MeV/fm$^{\rm 3}$.  We observe that the pressure increases as a function of energy density almost monotonically for KDE0v1, NRAPR and HV, whereas QMC700 EoS predicts a change in the rate of increase at about 4 $\epsilon_0$. This change, and the subsequent softening of the EoS, happens at the transition energy density marking the threshold for appearance of hyperons in the matter. Such a change is not observed in the HV EoS. The main reason for the difference between the two hyperonic models is that the QMC700 distinguishes between the nucleon-nucleon, and nucleon-hyperon interactions (neglecting the poorly known hyperon-hyperon interaction), whereas the HV model uses a universal set of parameters for all hadrons. Inclusion of both the QMC700 and HV EoS in this work reflects the uncertainty in the theory of dense matter in the cores of neutron stars.  

\section{Results and discussion}\label{sec:res}

As detailed in the previous section, the calculation of the frequency dependence of the braking index has been done for a multiple combination of codes, EoS's and M$_{\rm B}$ of the rotating star. We show only typical examples of the results, usually for the QMC700 EoS, unless stated otherwise.

\subsection{Braking index with frequency dependent MoI}
\begin{figure}[htb]
\includegraphics[width=0.6\textwidth]{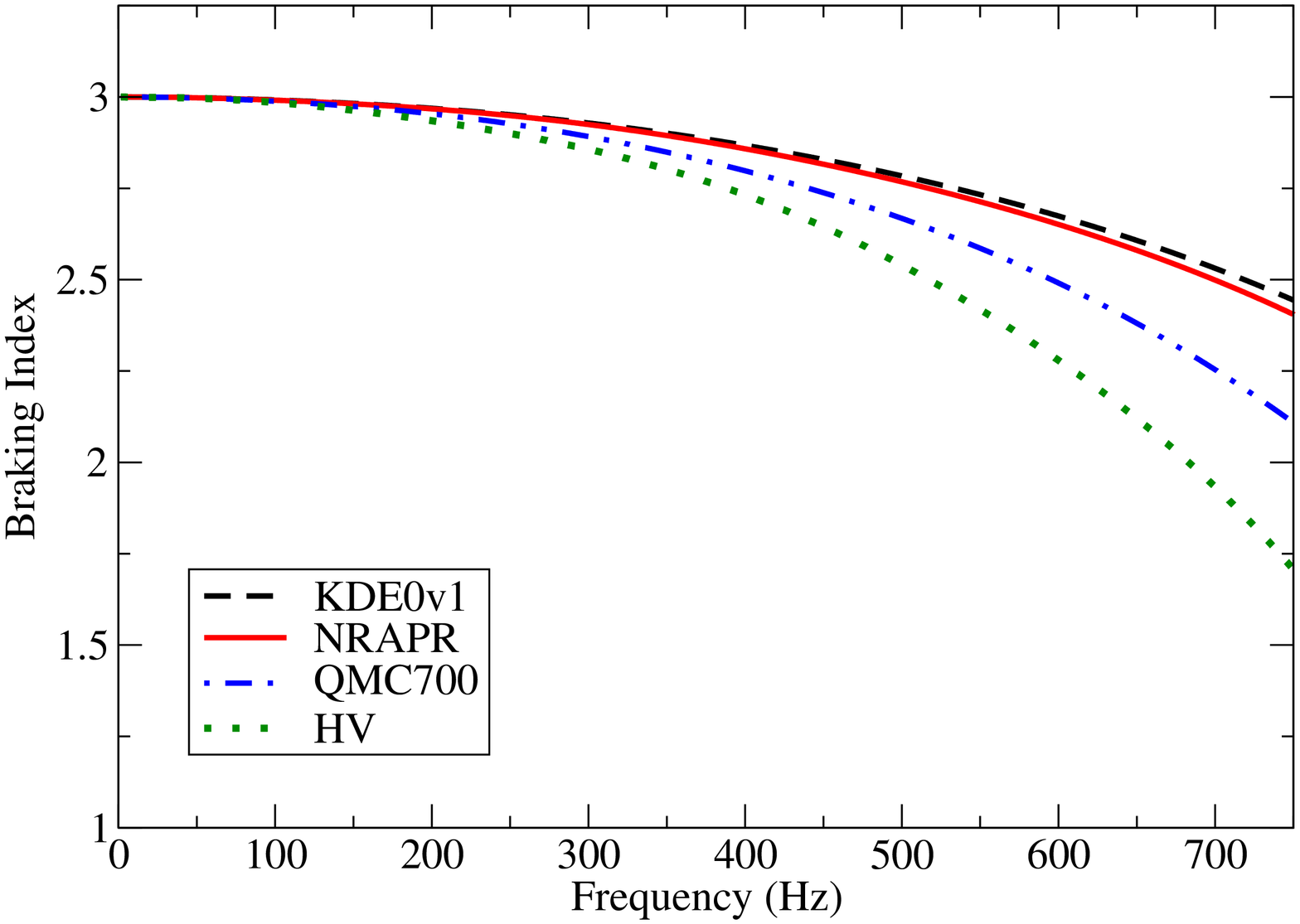}
\includegraphics[width=0.6\textwidth]{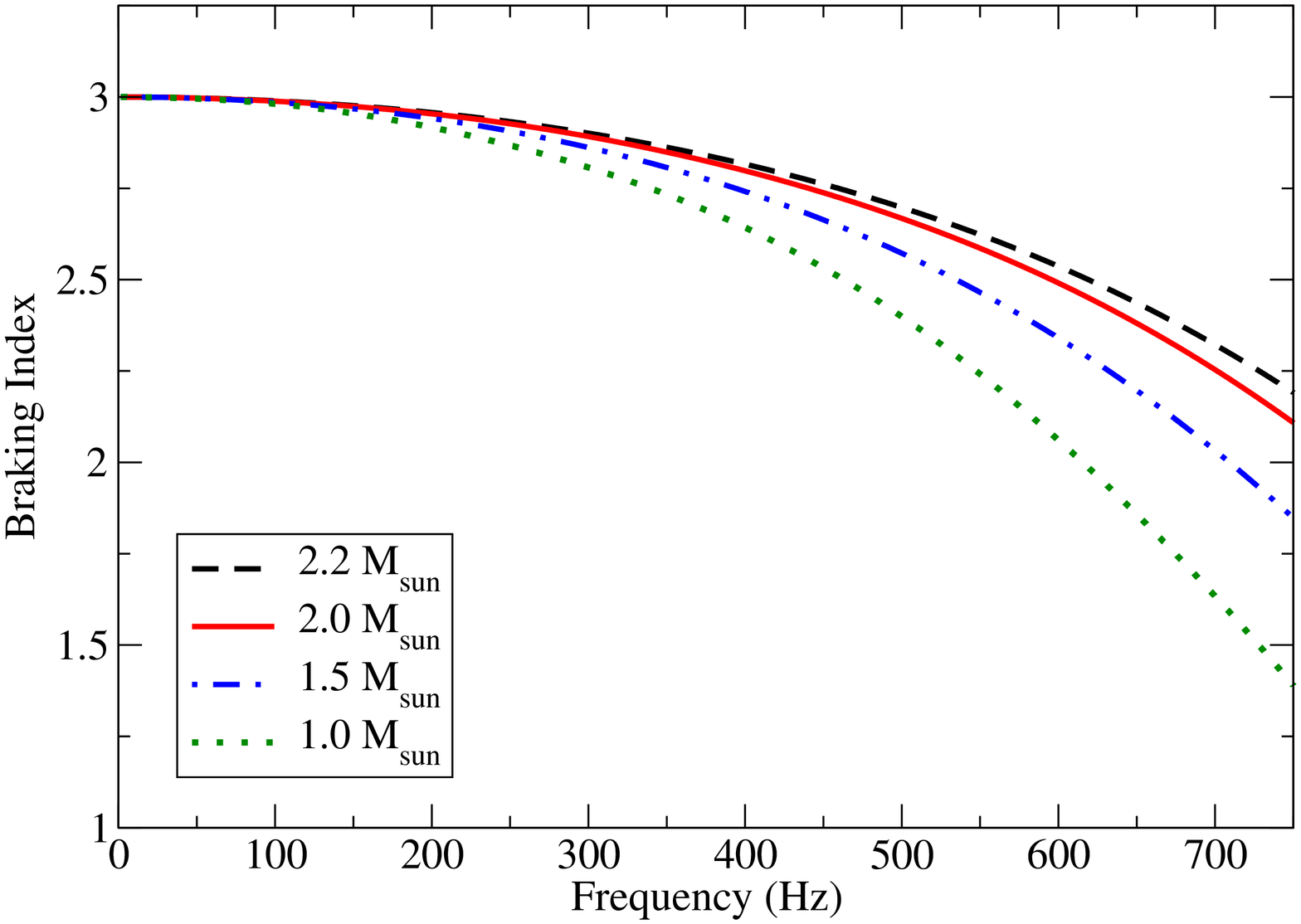}
 \caption{Braking index as a function of frequency.
 Left Panel: Braking index calculated for a pulsar with M$_{\rm B}$  = 2.0~$M_\odot$ with all EoS's adopted in this work.
Right Panel: Braking index calculated of pulsars with M$_{\rm B}$ =  1.0 -  2.2 M$_\odot$.}
\label{fig3}
\end{figure}

As a general feature, we find that any appreciable deviation of the braking index from the generic value $n = 3$ is observed only at rotational frequencies higher than about 250 Hz. The sensitivity of this deviation to the EoS and M$_{\rm B}$ is demonstrated in Figure~\ref{fig3}. As can be seen in  Figure~\ref{fig3},  the biggest change in the braking index of a 2.0 M$_\odot$ star pulsar is predicted by the HV EoS, followed by the QMC700, reaching $\sim$ values 1.75 and 2.15 at 750 Hz, respectively. The two nucleon-only EoS's, KDE0v1 and NRAPR behave in a very similar way and predict a larger value of n = 2.5 at this frequency. These trends can be directly related to the properties of the EoS's. Figure ~\ref{fig3} shows the sensitivity to  M$_{\rm B}$ for the QMC700 EoS. The effect clearly increases with decreasing M$_{\rm B}$.

\subsection{Superfluidity of the core}
The effects demonstrated in Figure~\ref{fig3} were calculated assuming that the whole body of a pulsar contributes to the total (core+crust) MoI. However, some theories suggest the conditions inside a pulsar are consistent with the presence of superfluid/superconducting matter, both in the crust and in the core \cite{sed1998, ho2012, page2014, hooker2013}. Superfluid material would not contribute to the rotation thus reducing the MoI. 

In this work we considered an extreme case in which the whole contribution of the core to the total MoI is removed. This scenario could be realized, for example, if either the whole core is superfluid or there is a layer of superfluid material between the core and the inner crust of the star, preventing an angular momentum transfer between the core and the crust. Either scenario simply results in removal of the contribution of the core to the MoI. 

\begin{figure}[htb]
\includegraphics[width=0.6\textwidth]{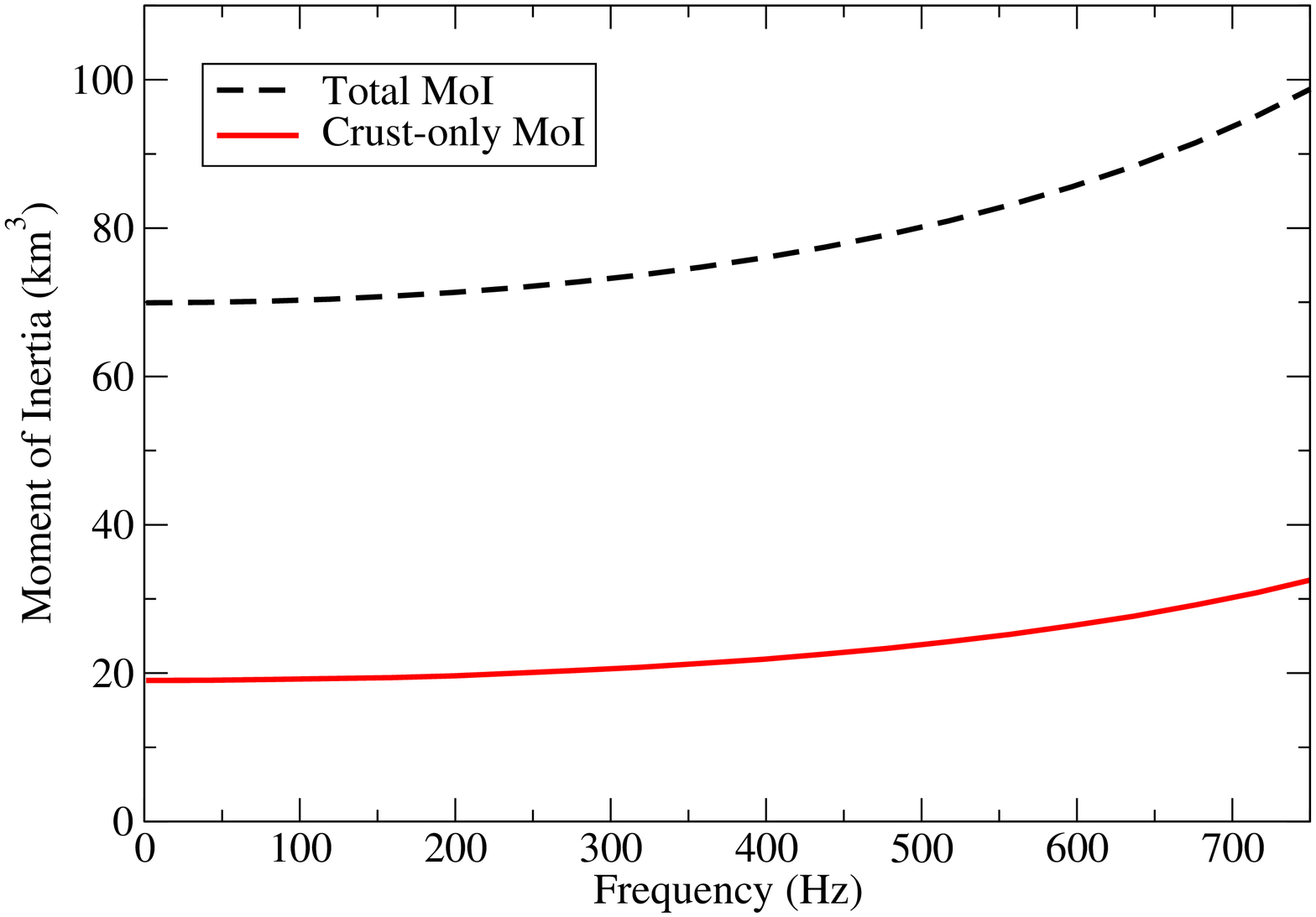}
\includegraphics[width=0.6\textwidth]{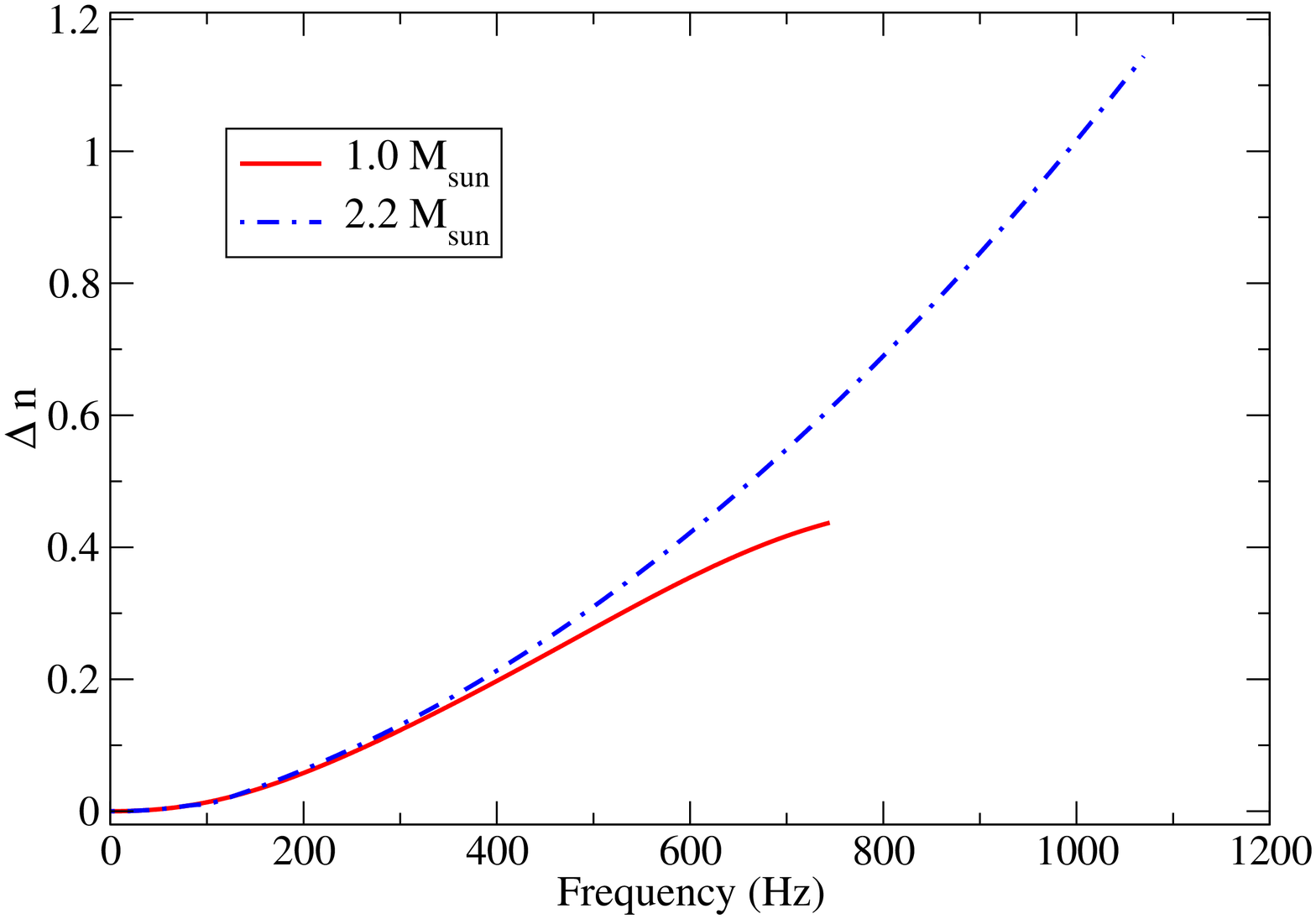}
\caption{Left panel: Total (crust and core) and the crust-only MoI as a function of frequency, calculated for a pulsar with M$_{\rm B}$  = 1.0 M$_{\odot}$.
Right panel: $\Delta$ n represent the difference in braking index as a function of frequency between stars with and without core contribution to the MoI.  Each curve is displayed up to the Kepler frequency of the star.}
\label{fig7}
\end{figure}

Elimination of the core contribution can lead to a dramatic lowering of the  total MoI by more than a factor of three, as shown, as an example, in Figure~\ref{fig7} for the 1.0~M$_\odot$ baryon mass and the QMC700 EoS. The difference between the total and crust-only MoI shows a weak frequency dependence with a slight increase above about 600 Hz. In turn, the reduction of the MoI by removal of the core contribution leads to additional changes in the braking index, on top of the changes due to the frequency dependent MoI (see Figure~\ref{fig3}) as shown in  Figure~\ref{fig7}. This change is, as expected, larger for higher mass stars which  contain a more significant proportion of dense core material then for lower mass stars being more crust-like throughout.

\subsection{Summary} 
The variation of  braking index of isolated rotating neutron stars with M$_{\rm B}$ = 1.0~M$_\odot$, 1.5~M$_\odot$, 2.0~M$_\odot$ and 2.2~M$_\odot$  with rotational frequency from zero to the Kepler limit within the MDR model with frequency dependent MoI has been investigated. The microphysics of the star was included through utilizing realistic EoS's of the pulsar matter. An illustration of the possible effect of superfluidity in the star core has been included in the study. 

\begin{figure}[htb]
\includegraphics[width=0.6\textwidth]{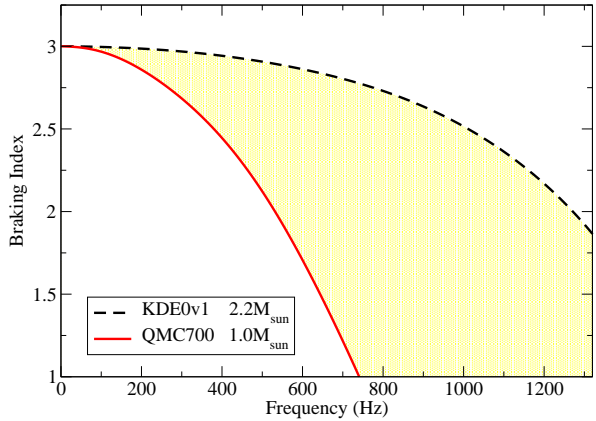}
\caption{Lower and upper limits on values of the braking index as a function of frequency, including results from both numerical codes, all EoS's, M$_{\rm B}$, and the superfluid condition. The (yellow) shaded area between the two lines defines the location of all results within the limits. The pulsar with baryonic M$_{\rm B}$ = 2.2  M$_{\odot}$ and the KDE0v1 EoS has the highest Kepler frequency and defines the frequency limit in this work.}
\label{fig9}
\end{figure}

Compiling results of all models used in this work, including the superfluidity effect, we deduce a definitive upper and lower limit on the braking index as a function of frequency, shown in Figure~\ref{fig9}.  The maximum change in the braking index is obtained with the QMC700 EoS and 1.0 M$_\odot$, the least effect is found for KDE0v1 and the 2.2 M$_\odot$ star.  Reduction of the braking index from the simple MDR model value $n = 3$ happens only at frequencies that are some significant fraction of the Kepler frequency. The calculation predicts that isolated pulsars with the braking index most deviating from $n = 3$ have low  M$_{\rm B}$. For the frequencies of known isolated pulsars with accurately measured braking indices (see Table~\ref{tab1}), the reduction away from $n=3$ found in this model, is negligible.

Finally, we have shown that the simple exclusion of the core due the superfluidity, or some superfluid  barrier between the crust and core, does not have a strong effect on braking in the frequency range of observed isolated pulsars. Further development of the idea of a macroscopic description of superfluidity would be interesting. Change of the magnetic field due to superfluidity and possible magnetic field expulsion, and a consequential increase in surface magnetic field strength $B$ could also be usefully explored.

%%%%%%%%%%%%%%%%%%%%%%%%%%%%%%%%%%%%%%%%%%%%%%%%%%%%%%%%%%%%%%%%%%%%%%%%%
%%
%%   use this format to include an .eps figure into your paper
%%
%\begin{figure}[htb]
%\includegraphics[width=0.6\textwidth]{Delta_Pauli}
%\includegraphics[width=0.6\textwidth]{Mott-Density}
%\caption{Binding energy $E_{B}$ and continuum edge $u(p=0,P_{F})$ for nucleons
%in symmetric nuclear matter (solid lines) and pure neutron matter (dashed
%lines) as a function of the density at $T=0$. 
%The nucleon dissociation (Mott transition) occurs at the densities where the 
%binding energy merges the corresponding continuum edge.
%Left panel: The role of the Pauli blocking for the Mott effect is shown.
%Right panel: The Mott densities are almost independent of the choice of the
%potential parameter $V_0$, see also Table \ref{tab:parameters}.}
%\label{fig:Mott}
%\end{figure}
%%%%%%%%%%%%%%%%%%%%%%%%%%%%%%%%%%%%%%%%%%%%%%%%%%%%%%%%%%%%%%%%%%%%%%%%%%%

\bigskip
\subsection*{Acknowledgement}
We express our thanks to the organizers of the CSQCD IV conference for providing an 
excellent atmosphere which was the basis for inspiring discussions with all participants.  We would like to acknowledge that this work was supported in part by Czech Grant No. GACR 209/12/P740 and Grant No. CZ.1.07/2.3.00/20.0071 "Synergy," aimed to foster international collaboration of the Institute of Physics of the Silesian University, Opava.  The research was also supported by the Department of Physics and Astronomy, University of Tennessee.

\end{document}